\documentclass[aps,prl,twocolumn,showpacs,preprintnumbers,amsmath,amssymb]{revtex4}
\usepackage{pifont}

\usepackage{graphicx}
\usepackage{dcolumn}
\usepackage{bm}
\usepackage{color}

\begin{document}

\preprint{APS/123-QED}

\title{Effective leadership in competition}

\author{Hai-Tao Zhang$^{1,2}$}
\author{Ning Wang$^2$}
\author{Michael Z. Q. Chen$^{3}$}
\author{Tao Zhou$^{4,5}$}
\author{Changsong Zhou$^{6,\dagger}$}
\affiliation{$^1$Department of Engineering, University of Cambridge, Cambridge CB2 1PZ, U.K. \\
$^2$Department of Control Science and Engineering, Huazhong
University of Science and Technology, Wuhan 430074, P.R. China\\
 $^3$Department of Automation, Nanjing University of Science and Technology,
Nanjing 210094, P.R. China\\
$^4$Department of Modern Physics, University of Science and Technology of China, Hefei 230026, P.R. China\\
$^5$Department of Physics, University of Fribourg, Chemin du Muse 3,
Fribourg CH-1700, Switzerland,\\
$^6$Department of Physics, Centre for Nonlinear Studies,
and The Beijing-Hong Kong-Singapore Joint Centre for Nonlinear and Complex Systems (Hong Kong),
Hong Kong Baptist University, Kowloon Tong, Hong Kong, China }

\begin{abstract}
Among natural biological flocks/swarms or even mass social
activities, when the collective behaviors of the followers has been
dominated by the moving direction or opinion of one leader group, it
seems very difficult for later-coming leaders to reverse the orientation of the mass
followers, especially when they are in
quantitative minority. This Letter  reports a counter-intuitive
phenomenon, \emph{Following the Later-coming
Minority},  provided that the late-comers obey a favorable
distribution pattern which enables them to spread  their influence
to as many followers as possible in a given time and to accumulate
enough power to govern these followers.
We introduce a discriminant index to quantify  the whole group's orientation under
competing  leadership, which helps to design an economic way for
the minority later-coming leaders to defeat the dominating majority
leaders solely by optimizing their distribution pattern. Our
investigation provides new insights into the effective leadership
in biological systems, with meaningful  implication to social
and industrial applications.
\end{abstract}

\maketitle
For biological flocks/swarms, their collective behavior always
depends on social interactions among group members. In many cases,
just a few individuals have the pertinent global information, like
the knowledge about the location of a food source or an obstacle, or  a migration
route.  It is known that several
species can evolve specific signals that help guide the uninformed
individuals \cite{co03a}. On the other hand, valuable leadership may
be correlated with age, status or reputation, and it is very common
for many species that experienced group members play an important
role in helping the less experienced. It was demonstrated that a
small proportion of informed individuals are sufficient to guide the
navigating behavior of the whole group, e.g., foraging fish schools
and bee swarms heading for new nest sites \cite{co03,co05}.

Nevertheless, the nature of bio-groups is not always that simple,
since it often happens that the informed individuals within a group
may differ with each other in their preferred directions due to
different experiences or motivations. This divergence also happens
frequently in human society, e.g., different political parties can
possess totally different beliefs. Interestingly, it is often
encountered that, when the mass followers' orientation has been
completely dominated by one leader group, another leader group, who
may be in quantitative minority, enters aiming at reverse the
followers' orientation to its own. For instance, migrating birds or
foraging insects developed an effective way to deviate to a new
promising direction at a low cost of additional leadership
\cite{co05}. In elections, a new social party always desires to
defeat its elder opponents with as few extra seats in a legislature
as possible. In marketing competition, the newly coming corporations
would manage to acquire more share from the market dominated by the
monopolies at a low cost.

In the last years, this  issue of effective leadership has
attracted significant attention \cite{co03,co05,su06,bu06}. In
general, many existing works show that the whole group is more
likely to follow the majority rather than the minority under the
guidance of divergent leadership \cite{br96}.
For instance, the cost to the groups as a whole
is considerably higher for a ``despotic" than for a ``democratic
decision" or ``following the majority" manner \cite{co03}.
The larger the group,
the smaller the proportion of informed individuals is needed to
guide the group \cite{co05}.
In social science, it is shown that the
public is apt to follow the majority under the impression of
divergent opinions from different social parties \cite{su08, an04}.
The synchronization of many hands clapping \cite{na00} and the
escaping panic \cite{he00} strongly imply the rule of
``following the majority". All these bring up an important question:
 \emph{is it possible for the minority later-coming leaders
defeat the dominating majority ones and how}?

In this Letter we address this question in a generic model of
collective behavior,  the Vicsek model \cite{vi95},
in which individuals align motions to the average of their
geographical neighbors to achieve the global velocity
synchronization. Importantly, we have found
that   defeating by minority  is highly possible provided that the
later-comers adopt better distribution pattern to
influence and persuade more followers. To quantify the effectiveness
of the leadership, we propose an evaluation index which can
 predict reasonably the orientation of the mass followers under the
guidance of diverged leadership solely  basing on the parameters of the distribution patterns.
%


The original Vicsek model is extended to incorporate the influence of the earlier and later-coming
leaders:  a small proportion of the whole group of $N$-individuals is
given a preferred motion direction representing, for example, the
direction to a known food resource or a migration target, or the
faithfulness for one political belief or commodity brand.
In our model there are three types of
individuals: i) $N_r$ earlier-coming leaders moving rightwards,
namely, $\mathcal{N}_r$, whose dynamics are
$x_{r_i}(t+1)=x_{r_i}(t)+v\measuredangle 0^o, ~i=1,\cdots, N_r$; ii)
$N_l$ minority later-coming leaders moving leftwards, namely,
$\mathcal{N}_l$, whose dynamics are
$x_{l_i}(t+1)=x_{l_i}(t)+v\measuredangle 180^o, ~i=1,\cdots, N_l$
with $N_l\leq N_r$; and iii) $N_f$ uninformed individuals, namely,
$\mathcal{N}_f$, whose dynamics are updated by $x_{f_i}
(t+1)=x_{f_i}(t)+v\measuredangle \theta_{f_i}(t)$, $i=1,\cdots,N_f$,
with $N_f= N-N_r-N_l$. Here, $x_i$ denotes the position of
individual $i$, and the leaders move without being  affected  the
others. $\mathcal{N}_f$ are naive and have no preference  in
any particular direction but just follow the average directions
$\theta_{f_i}(t)$ of their neighbors, and cannot differentiate the
leaders and the followers. Note that, at the beginning, there are
just $\mathcal{N}_r$ and $\mathcal{N}_f$ and  once the
orientation of $\mathcal{N}_f$ has completely aligned to
$\mathcal{N}_r$, $\mathcal{N}_l$ appears to compete with
$\mathcal{N}_r$ to reverse  the orientation of $\mathcal{N}_f$.

The velocity of the $f_i$-th follower, i.e. $v_{f_i} (t)$, has a
constant speed $v$ and a direction $ \theta
(t+1)=\left<\theta\left(t\right)\right>_r$, where
$\left<\theta\left(t\right)\right>_r$ denotes the average direction
of individuals within a circle of radius $r$ surrounding individual
$i$ (including itself) \cite{vi95}.
This kind of aligning mechanism can
nicely mimic the local dynamics of ``\emph{go
with the stream}" in both bio-groups and human society.  To
focus on the effects of the leaders, we do not consider the influence of
the external noise.

Here, without loss of generality, we set $N=500$ and  $L=10$ with
periodic boundary conditions, and $r=1$, $v=0.03$ as Ref.
\cite{vi95}.
The global orientation of the whole group is defined as  normalized steady-state
alignment index $V_m=1-\theta_a/(\pi /2)$, where $\theta_a$ denotes
 the steady-state direction of the whole group, thus the values $1$,
$-1$ and $0$ of $V_m$  mean that  the whole group is completely following
$\mathcal{N}_r$, $\mathcal{N}_l$ and no preference in direction,
respectively.

\begin{figure}[htp]
\centering
  \resizebox{5.0cm}{!}{\includegraphics[width=12.0cm]{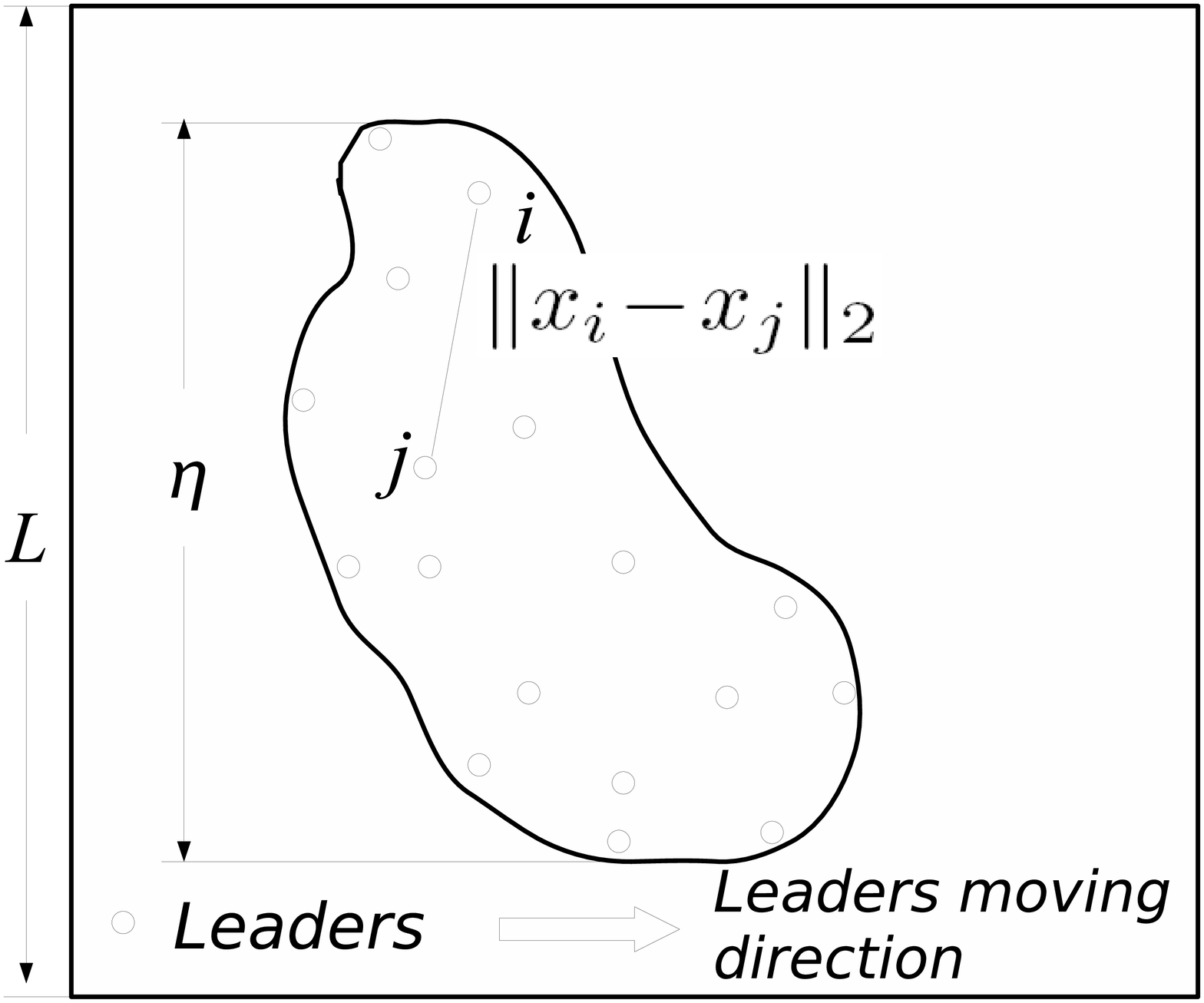}}
\caption{Illustration of effective leadership factors
\emph{\textbf{F1}} and \emph{\textbf{F2}} by normal length $\eta$
and clustering factor $\sigma$, whose definitions will be given
later. Here, $\| x \|_{2}=\sqrt{x^{T}x}$.}
 \label{fig: factors}
\end{figure}

Now recall the key problem this Letter addresses: \emph{is it
possible for $\mathcal{N}_l$ to defeat $\mathcal{N}_r$?}
Questions closely relevant to this issue have already kindled up the
interests of not only physicists and biologists but also social
scientists and marketing researchers for years \cite{co05,su08}. In
these studies, it is generally drawn that the followers would be apt
to follow the majority leaders since in the alignment models
 each individual follows the average direction of its
neighbors. Then, it can be  seen that, if the majority and the
minority have the same geographical distribution pattern, the
majority has either larger influence area or higher particle density
which intensifies their leadership. Thereby, it is  natural to
deduce that decisive parameter(s) for effective leadership may be
not the absolute number of the leaders  but the following two key,
but competing factors: \emph{\textbf{F1}}) \emph{Effective Range}:
to distribute leaders' influence to as many followers as possible
within a given time; \emph{\textbf{F2)}} \emph{Persuasive
Intensity}: to be sufficiently persuasive to govern the followers
they can influence with high density. In other words, minority leaders may defeat the
majority ones provided that they have better \emph{\textbf{F1}} or
\emph{\textbf{F2}} or both, and our work will verify such a
hypothesis. In our scenario given above, \emph{\textbf{F1}} and
\emph{\textbf{F2}} can be quantified by the normal length $\eta$ and
clustering factor $\sigma$, as shown in Fig.~\ref{fig: factors}.
Obviously, these two factors are however somewhat contradictive
since \emph{\textbf{F1}} requires the leaders to distributed
sparsely into the followers' region while \emph{\textbf{F2}} favors
highly condensed leader groups, and this contradiction constitutes
the main challenge  of the problem.

\begin{figure}[htp]
\centering
  \resizebox{8.3cm}{!}{\includegraphics[width=5.0cm]{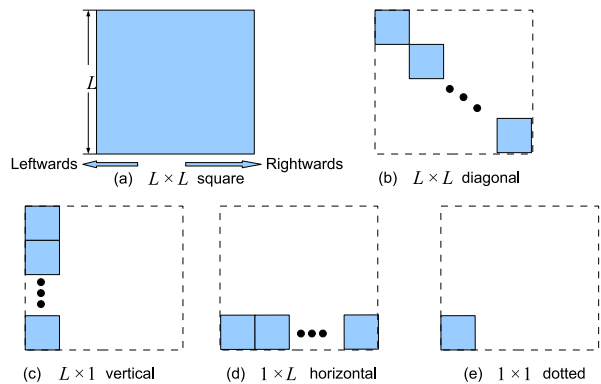}}
\caption{(Color online) Typical distribution patterns of leaders. }
 \label{fig: pattern}
\end{figure}

To address this interesting question, we have implemented
simulations with $\mathcal{N}_r$ distributed in the $L\times L$
square (Fig.~\ref{fig: pattern}(a)), as considered in most previous works.
We then examine the effectiveness of $\mathcal{N}_l$ leader group  on
various distribution patterns,  such as the $L\times L$ square, $L\times L$ diagonal,
$L\times 1$ vertical, $1\times L$ horizontal and $1\times 1$ dotted
regions as shown in Figs.~\ref{fig: pattern}(a)--(e), respectively.
According to the two factors \emph{\textbf{F1}} and
\emph{\textbf{F2}} of effective leadership and taking into
consideration of the moving directions of $\mathcal{N}_l$ and
$\mathcal{N}_r$, one can expect that $L\times 1$ outperforms
$L\times L$ because they have the identical \emph{\textbf{F1}} but
$L\times 1$ favors \emph{\textbf{F2}}. Analogously, $L\times 1$ can
also be expected to be superior to  $1\times L$ since they have the same
\emph{\textbf{F2}} but $L\times 1$ have better \emph{\textbf{F1}}.
However, it is difficult to compare $L\times 1$ and $1\times 1$,
since $L\times 1$ has better \emph{\textbf{F1}} while $1\times 1$
greatly favors \emph{\textbf{F2}}. Thus, one has to resort to
numerical simulations to reveal more concrete rules behind.


\begin{figure}[htp]
\centering
\resizebox{7.0cm}{!}{\includegraphics[width=12.0cm]{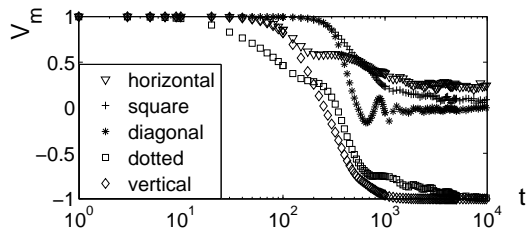}}
\vspace*{-2.1cm} \caption{Orientation reversion of the mass
followers under the later-coming leaders $\mathcal{N}_l$ obeying 5
typical different distribution patterns given in Fig.~\ref{fig:
pattern}. The follower group $\mathcal{N}_f$  has completely aligned
to the leaders $\mathcal{N}_r$ when $\mathcal{N}_l$ sets in at
$t=0$.
 Here, $N_r=N_l=10$.
Each point is an average over $1000$ independent
runs for this and the following figures.}
 \label{fig: distribution}
\end{figure}
Four more interesting and concrete phenomena are
observed from Fig.~\ref{fig: distribution}: i) $L\times 1$ vertical
and $1\times 1$ dotted patterns (Figs.~\ref{fig: pattern}(c) and
(e)) can defeat the earlier-coming $L\times L$ square pattern and
reverse the orientation of the followers, while $1\times L$
horizontal, $L\times L$ square and $L\times L$ diagonal patterns
(Figs.~\ref{fig: pattern}(a), (b) and (d)) not. ii) $L\times 1$
vertical and $1\times L$ horizontal patterns are the most and least
effective ones, respectively; iii) $L\times L$ diagonal distribution
is a little bit better than the $L\times L$ square distribution; iv)
$1\times 1$ dotted pattern is the second most effective one (just
below $L\times 1$). Inspiringly, for $\mathcal{N}_l$ adopting a
dotted  distribution $1\times 1$, it takes
considerable running steps to propagate its influence to remote
followers, so that the converging time is much longer.
These  simulation results help us understand more deeply
the nature of \emph{\textbf{F1}}and \emph{\textbf{F2}}.
Specifically, in our scenario,\emph{\textbf{F1}} has been realized
by spreading $\mathcal{N}_l$ out sufficiently \emph{perpendicularly} to
their movement direction, while a condensed distribution corresponding to
 favorable \textbf{\emph{F2}} could first slave the followers locally and then
 propagate the influence to the whole  population.

\begin{figure}[htp]
\centering
\begin{tabular}{cc}
  \hspace*{-0.3cm}
  \resizebox{4.8cm}{!}{\includegraphics[width=12.0cm]{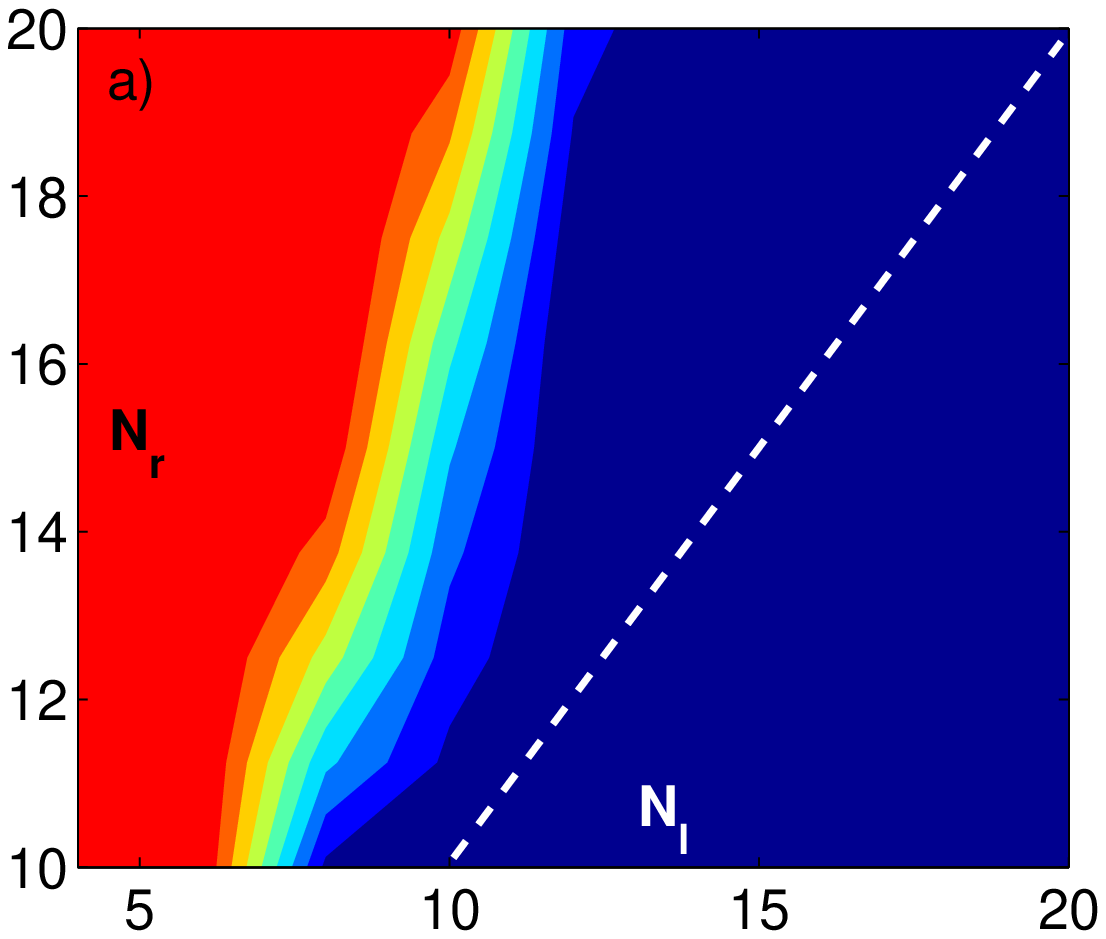}}
  & \hspace*{-0.8cm}
  \resizebox{4.8cm}{!}{\includegraphics[width=12.0cm]{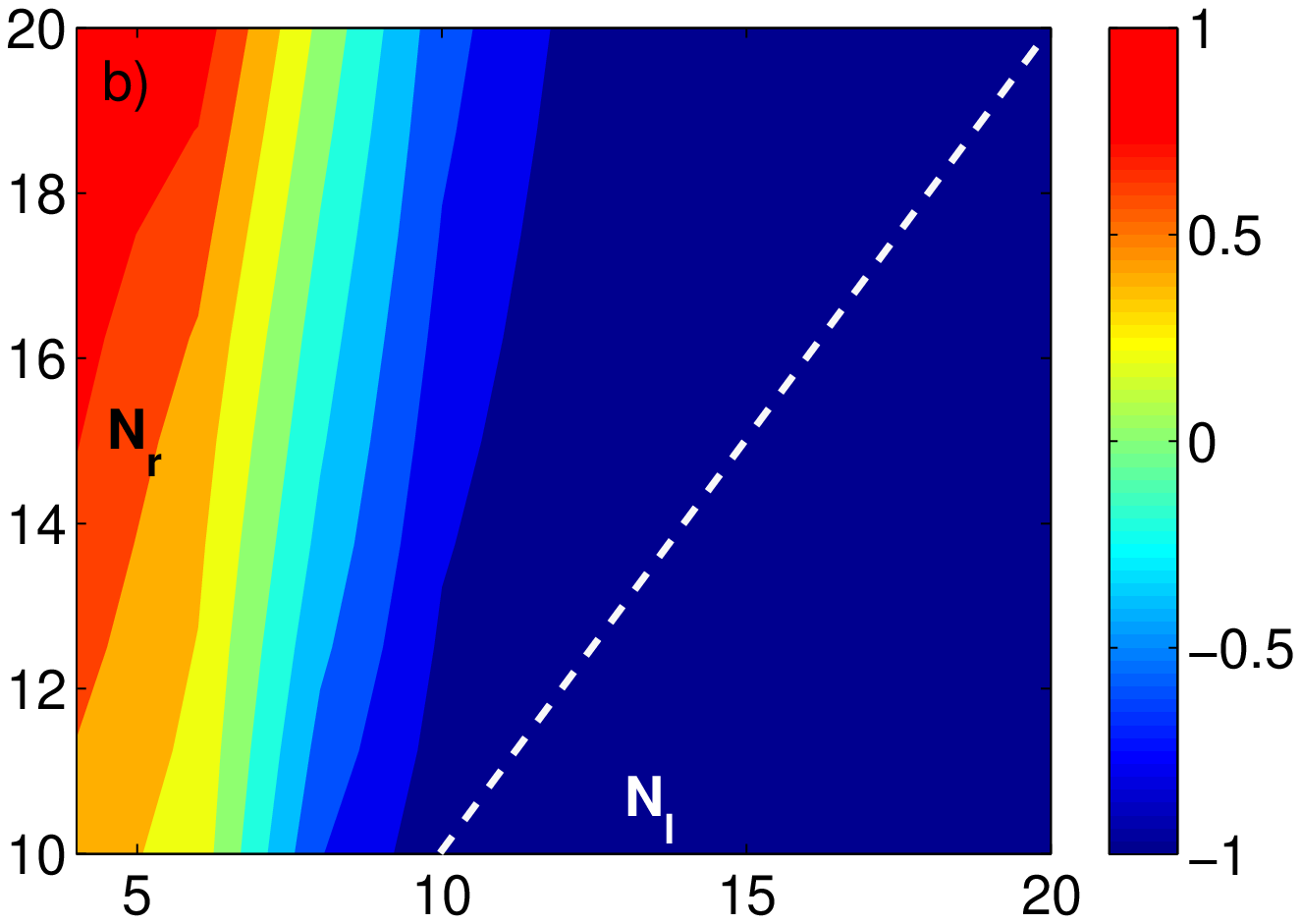}}
\end{tabular}
\vspace*{-0.6cm}\caption{(Color online) Effective leadership for $\mathcal{N}_l$ vs
$\mathcal{N}_r$ leaders with different distribution patterns  $L\times 1$ vertical  and
$L\times L$ square, respectively.
(a) $V_m$ from the model simulations. (b) $V_{m1}$ from Eq. (1).
}
 \label{fig: following the majority or minority}
\end{figure}

To this end, one can be delighted to infer that even if
$\mathcal{N}_f$ have been completely dominated by $\mathcal{N}_r$,
it is highly possible for the minority $\mathcal{N}_l$ to reverse
the followers' opinion with their better distribution pattern. Indeed, as
demonstrated clearly in Fig.~\ref{fig: following the majority or minority} (a),
in a large region above the white dashed line where $N_l<N_r$, the whole
group reverses the orientation from $V_m=1$ to $V_m=-1$ (blue region) to
 follow the leadership of the minority  $\mathcal{N}_l$ which has  a better distribution pattern
 ($L\times 1$ vertical) compared to the majority ($L \times L$ square).
 Similar results are observed for the other favorable $1 \times 1$ dot pattern of $N_l$.


The numerical simulations suggest that  the two factors
\textbf{\emph{F1}} (influencing area) and \emph{\textbf{F2}}
(clustering intensity) determining the leadership performance can be
quantified as below. As shown in Fig.~\ref{fig:
factors}, \emph{\textbf{F1}} can be represented by the length  of
the leaders' distribution region perpendicular to the movement
direction of the leader, namely the \emph{normal length} $\eta$. The
clustering intensity \emph{\textbf{F2}} can be characterized by the reciprocal of the average geographical distance among  the leaders,
namely the \emph{clustering factor} $\sigma$.
We find that for the number of leaders $N_l$,
the average geographical distance between the first $2N_l $ nearest pairs of leaders
can   sensitively distinguish  various patterns discussed in Fig. 2 and Fig. 5.
More precisely, for the $N_l$ leaders,
$\sigma_{l}=1/(\frac{1}{2N_{l}}\sum_{i,j\in \mathcal{N}_{l},~j\neq
i,\|x_i-x_j\|_{2} \leq\bar{d}_{2N_{l}}}\|x_i-x_j\|_{2})$. Here,
$\bar{d}_{2N_{l}}$ denotes the geographical distance between the
$2N_{l}$-th nearest pair of the leader group. The same formula  hold
for $\sigma_r$ of  the leader group $\mathcal{N}_r$.


\begin{figure}[htp]
\centering
\begin{tabular}{cc}
  \hspace*{-0.3cm}
  \resizebox{4.5cm}{!}{\includegraphics[width=10.0cm]{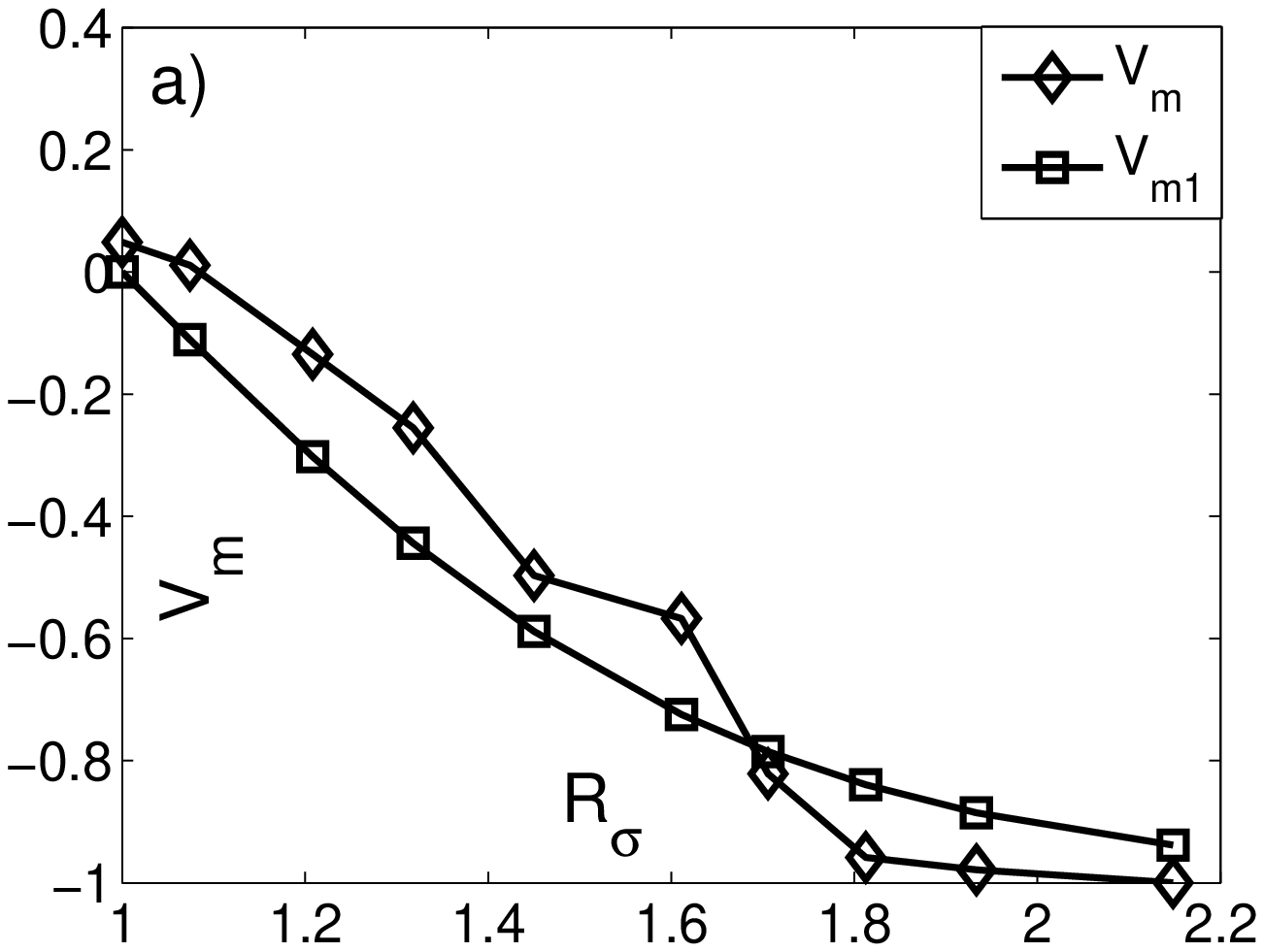}}
  & \hspace*{-0.2cm}
  \resizebox{4.5cm}{!}{\includegraphics[width=10.0cm]{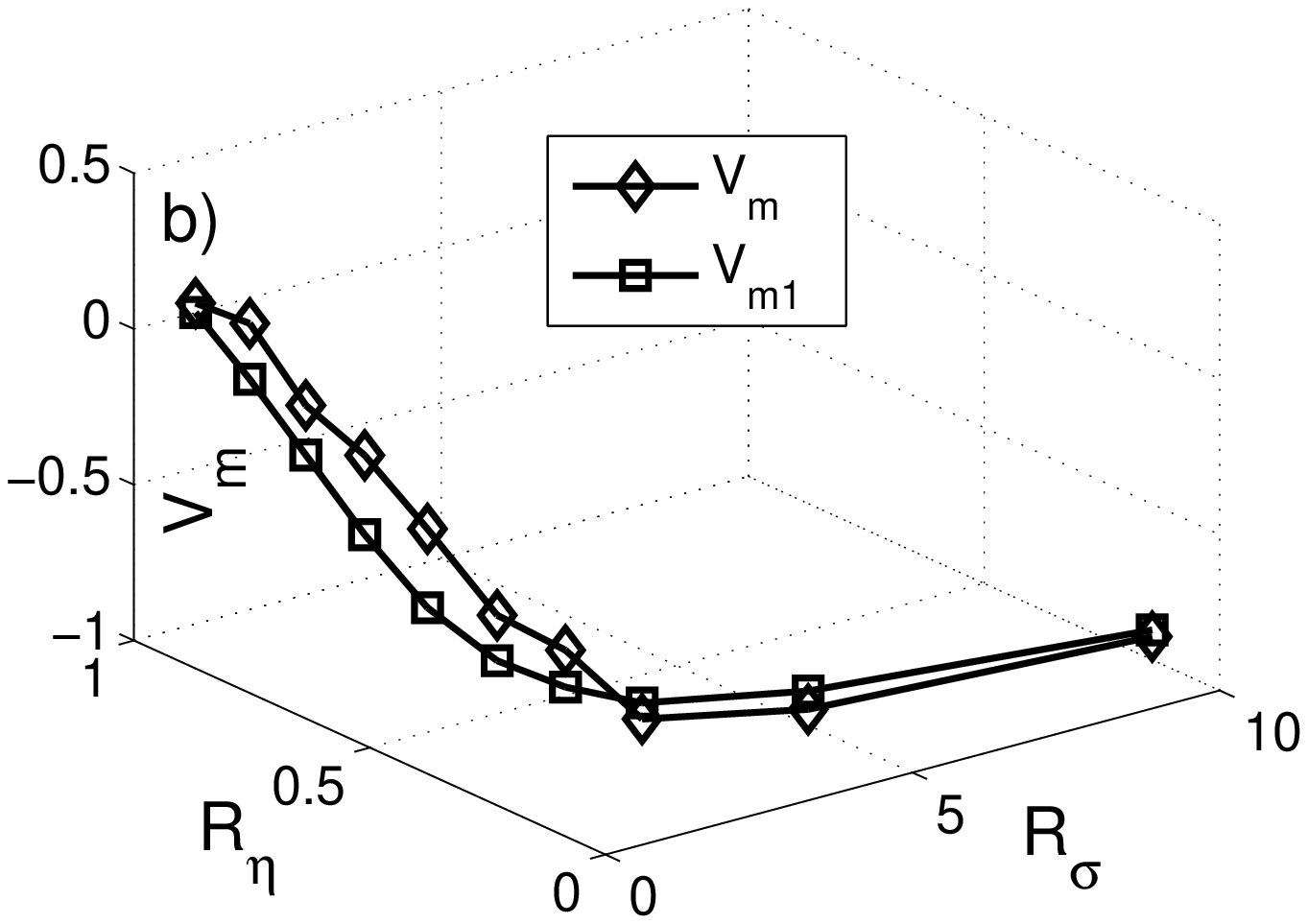}}
\end{tabular}
\caption{Roles of clustering factor ratio $R_{\sigma}$ and normal
length ratio $R_{\eta}$  on the orientation of the followers. (a)
Effect of $R_{\sigma}$; (b) Combined effects of both $R_{\sigma}$
and $R_{\eta}$. $V_{m1}$ in Eq. (1) is compared to $V_m$ from the
Vicsek model simulations. }

 \label{five}
\end{figure}

Now with these factors $\eta_{l,r}$  and $\sigma_{l,r}$ (normal lengths and clustering factors of
$\mathcal{N}_l$ and $\mathcal{N}_r$, respectively), one is ready to make a concrete comparison
between different leaderships. To quantify the effective
leadership using these factors,  we further  define an influencing region
ratio $R_{\eta}=\eta_l/\eta_r$ and a clustering intensity ratio
$R_{\sigma}=\sigma_l/\sigma_r$.
Remarkably, we find that the effective leadership can be
reasonable predicted solely by these parameters of the leader groups.
We let $\mathcal{N}_r$ randomly
distribute in the unbiased $L\times L$ square (Fig.~\ref{fig:
pattern}(a)), and let $\mathcal{N}_l$ (with
$N_l=N_r=10$) randomly distribute in $L\times L$, $L\times (L-1)$,
$\cdots$, $L\times 1$ rectangular regions, respectively, and hence
$\eta_r=\eta_l=L$ ( $R_{\eta}=1$) while $\sigma_l$ is increasing
monotonously. As shown in Fig.~\ref{five}(a), $V_m$ drops with
increasing $R_{\sigma}$ until asymptotically approaching a
saturation value of $-1$, indicating  dominant leadership of $\mathcal{N}_l$.
To investigate the
role of $R_{\eta}$ and $R_{\sigma}$ simultaneously,
we let $\mathcal{N}_l$ (with $N_l=N_r=10$) randomly
distribute in $L\times L$, $(L-1)\times (L-1)$, $\cdots$, $1\times
1$ square regions, respectively, which implies that $\eta_l$ is falling
whilst $\sigma_l$ is rising along this distribution sequence.
 Fig.~\ref{five}(b) shows  that $V_m$
reduces with increasing $R_{\eta}$ and $R_{\sigma}$.
As a consequence, one can confidently draw
that the minority later-coming leaders do have the potential to
reverse the followers if only they have larger value of $\eta$ or
$\sigma$ or both.

The observation from these intensive simulations  suggest that
we could  determine effective leadership for the two competing leader groups
solely basing on specific  combinations of the geometrical parameters
$\eta$ and $\sigma$. In fact, taking into consideration of the maximal and minimal saturation
values $1$ and $-1$ of $V_m$ and  $V_m(1,1)=0$, we hereby
propose a discriminant index $V_{m1}(R_{\eta},R_{\sigma})$ by
\begin{equation}\label{eq: index new}
V_{m1}(R_{\eta},R_{\sigma}) =w_1 \tanh[\gamma(1-R_{\sigma})]+w_2
\tanh(1-R_{\eta}).
\end{equation}
Here, $\gamma$ is used to adjust the origin-traversing slope of
$\tanh(\cdot)$ function, which endows $V_{m1}$ an essential degree
of freedom. According to our extensive numerical simulations in Fig. 5,
$\gamma\in [1.3,1.6]$ yields satisfactory approximation
performance. Thereby, without loss of generality, we set
$\gamma=1.5$, and then apply Least Square Estimation to identify
$w_1=1.0$, $w_2=0.2$ commonly for all the results in Fig. 5.
Note that this index is self-consistent in
the sense that: i) the maximal and minimal saturation values keep at
$1$ and $-1$ for the feasible ranges of $R_{\sigma}$ and $R_{\eta}$,
and ii) if either value of $R_{\sigma}$ and $R_{\eta}$ is 1 then
$V_{m1}$ will be merely determined by the other one.

It is important to note that  the definition of the clustering factor $\sigma$ naturally
takes the effect of the number of leaders $N_l$ and $N_r$ into account.
Remarkably,  Eq. (1) with the same parameter $\gamma=1.5, w_1=1.0$ and $w_2=0.2$
can account for the effective leadership for fixed patterns, but varying numbers
$N_l$ and $N_r$ ( Fig.~\ref{fig: following the majority or
minority}(b)). A comparison of the  Figs.~\ref{fig: following the majority or
minority} (a) and (b) shows clearly that $V_{m1}$ basing on the geometrical characterization
of the distribution patterns of the leaders can very nicely predict the effective
leadership $V_m$ in the model .

In summary, uncovering the nature of effective leadership is of great theoretical
and practical significance.
We have shown that  later-coming leaders,
even in quantitative minority, have the potential to defeat the
earlier-coming dominating ones, if only the former obeys a better
distribution pattern.   A better distribution pattern has larger
influential region and greater clustering factor, which can equip
the leaders with the capability of influencing more followers in a
given period and strengthening the persuasion power on the followers
as well. Intriguing enough, the mechanism underlying
such an apparent ``following the minority''  in the whole
group  is due to  the  scheme of ``following the majority'' locally.
Moreover, we have demonstrated that an index merely basing on the
geometrical parameters of the distribution patterns of the leaders  can provide
nice prediction of the effective leadership in competition.
With this index
one can quantify the
advantage of one leader group over another
so as to design an economical
way for the later-coming leaders to defeat the majority
earlier-coming ones.
Our simulations  on the other two more sophisticated models,
 the   Couzin's three-sphere model \cite{co03a} and the alignment
model \cite{co05},   strongly support  our conclusion on
the effective leadership mechanism.

Our investigation has launched a new exploration on the essential rules
that govern leadership potentials.
Motivated by both natural and social systems,
our findings have potential industrial and social
applications as well. The results are  valuable  to explain how the migrating
birds or foraging insects deviate to a new promising direction at a
low cost of additional leadership. The findings  are  also helpful to endow the
newborn corporations with some economic strategy to compete with
their dominating opponents in marketing competitions. Moreover,
industrial multi-agent systems can be expected to benefit from this
work to improve its adaptability to a new environment.
%

The work is  partially  supported by the  National Natural Science
Foundation of China (NNSFC, Grant No. 60704041) and the
Research Fund for the Doctoral Program of Higher Education (RFDP, Grant No. 20070487090) (HTZ), the NNSFC (Grant No. 10635040) (TZ) and
by the Hong Kong Baptist University (CSZ).


\begin{thebibliography}{100}
\bibitem{co03a} I.D. Couzin, J. Krause, R. James, G.D. Ruxton and N.R. Franks, J. Theor. Biol. \textbf{218}, 1 (2002).
\bibitem{co03} L. Conradt and T. J. Roper, Nature \textbf{421}, 155 (2003).
\bibitem{co05} I.D. Couzin, J. Krause, N.R. Franks and S.A. Levin, Nature \textbf{433}, 513 (2005).
\bibitem{su06}D.J.T. Sumpter, Philosophical Transactions of the Royal Society
B \textbf{361}, 5 (2003).
\bibitem{bu06} J. Buhl, D.J.T. Sumpter, I.D. Couzin, J.J. Hale, E. Despland,
E. R. Miller and S. J. Simpson, Science \textbf{312}, 1402 (2006).
\bibitem{br96} C.K.W. De Dreu and N.K. De Vries, British Journal of Social
Psychology \textbf{35}, 77 (1996).
\bibitem{su08} S. Suo and Y. Chen, Journal of Artificial Societies
and Social Simulations \textbf{11}, 42 (2008).
\bibitem{an04}M. Anghel, Z. Toroczkai, K. E. Bassler and G Korniss,
Phys. Rev. Lett. \textbf{92}, 058701 (2004).
\bibitem{na00} Z. Nada, E. Ravasz, Y. Brechet, T. Vicsek and A.-L. Barabasi, Nature \textbf{403}, 849 (2000).
\bibitem{he00} D. Helbing, I. Farkas and T. Vicsek, Nature
\textbf{407}, 487 (2000).
\bibitem{vi95} T. Vicsek, A. Czir\'ok, E. Ben-Jacob, I. Cohen, and O. Shochet, Phys. Rev. Lett.
\textbf{75}, 1226 (1995).




\end{thebibliography}

\end{document}